\pdfoutput=1

\documentclass[sigconf,nonacm]{acmart}

\usepackage{graphicx}

\usepackage{booktabs}

\usepackage{tabularx}
\usepackage{multirow}

\usepackage{subcaption}

\usepackage{seqsplit}

\usepackage{makecell}

\usepackage{verbatim}

\usepackage{amsmath}

\usepackage[nameinlink]{cleveref}

\usepackage{tabularray}

\usepackage{textcomp}

\usepackage{pifont}
\usepackage{xspace}
\usepackage{adjustbox}

\usepackage{xurl}

\usepackage{listings}
\definecolor{codegreen}{rgb}{0,0.6,0}
\definecolor{codegray}{rgb}{0.5,0.5,0.5}
\definecolor{codepurple}{rgb}{0.58,0,0.82}
\definecolor{backcolour}{rgb}{0.95,0.95,0.92}

\lstdefinestyle{mystyle}{
    backgroundcolor=\color{backcolour},   
    commentstyle=\color{codegreen},
    keywordstyle=\color{magenta},
    numberstyle=\tiny\color{codegray},
    stringstyle=\color{codepurple},
    basicstyle=\ttfamily\footnotesize,
    breakatwhitespace=false,         
    breaklines=true,                 
    captionpos=b,                    
    keepspaces=true,                 
    numbers=left,                
    numbersep=5pt,                  
    showspaces=false,                
    showstringspaces=false,
    showtabs=false,                  
    tabsize=2,
    frame=single,
    framesep=2pt,
    framexleftmargin=8pt,
    xleftmargin=0pt,
    xrightmargin=0pt
}

\newcommand{\filename}[1]{{\ttfamily\small\seqsplit{#1}}}

\newcommand{\totalnumtools}{eight}

\newcommand{\totalnumfunctions}{1,092,820}

\newcommand{\fscore}{F\textsubscript{1}-score}

\newcommand{\cmark}{\ding{51}}
\newcommand{\xmark}{\ding{55}}

\crefname{lstlisting}{listing}{listings}
\Crefname{lstlisting}{Listing}{Listings}

\newcommand{\eg}{e.g.,\xspace}
\newcommand{\ie}{i.e.,\xspace}
\newcommand{\etAl}{et\,al.\xspace}

\usepackage[framemethod=TikZ]{mdframed}
\global\mdfdefinestyle{insightstyle}{
	outerlinewidth=0.5pt,
	backgroundcolor=gray!10,
	linecolor=gray,
	innertopmargin=6pt,
	innerbottommargin=6pt,
	innerrightmargin=7pt,
	innerleftmargin=6pt
}
\global\mdfdefinestyle{recommendationstyle}{
	backgroundcolor=gray!10,
	linecolor=gray,
	innerrightmargin=7pt,
	innerleftmargin=6pt
}
\newenvironment{lesson}[1][Lessons learned.]
{
\smallskip
\begin{mdframed}[style=insightstyle]
\textbf{#1}
}
{
\end{mdframed}
}

\microtypecontext{spacing=nonfrench}
\graphicspath{{img/}}

\begin{document}

\title{Padding Matters -- Exploring Function Detection in PE Files}

\author{Raphael Springer}
\orcid{0009-0008-8298-7895}
\affiliation{
  \institution{Westphalian University of Applied Sciences}
  \department{Institute for Internet Security}
  \city{Gelsenkirchen}
  \state{}
  \country{Germany}
}
\email{springer@internet-sicherheit.de}

\author{Alexander Schmitz}
\orcid{0009-0006-0514-9535}
\affiliation{
  \institution{Westphalian University of Applied Sciences}
  \department{Institute for Internet Security}
  \city{Gelsenkirchen}
  \state{}
  \country{Germany}
}
\email{schmitz@internet-sicherheit.de}

\author{Artur Leinweber}
\orcid{0009-0001-7623-1038}
\affiliation{
  \institution{Westphalian University of Applied Sciences}
  \department{Institute for Internet Security}
  \city{Gelsenkirchen}
  \state{}
  \country{Germany}
}
\email{leinweber@internet-sicherheit.de}

\author{Tobias Urban}
\orcid{0000-0003-0908-0038}
\affiliation{
  \institution{Westphalian University of Applied Sciences}
  \department{Institute for Internet Security}
  \city{Gelsenkirchen}
  \state{}
  \country{Germany}
}
\email{urban@internet-sicherheit.de}

\author{Christian Dietrich}
\orcid{0009-0001-5523-4467}
\affiliation{
  \institution{Westphalian University of Applied Sciences}
  \department{Institute for Internet Security}
  \city{Gelsenkirchen}
  \state{}
  \country{Germany}
}
\email{dietrich@internet-sicherheit.de}

\begin{abstract}

Function detection is a well-known problem in binary analysis. 
While previous research has primarily focused on Linux/ELF, Windows/PE binaries have been overlooked or only partially considered.
This paper introduces FuncPEval, a new dataset for Windows x86 and x64 PE files, featuring Chromium and the Conti ransomware, along with ground truth data for \totalnumfunctions{} function starts.
Utilizing FuncPEval, we evaluate five heuristics-based (Ghidra, IDA, Nucleus, rev.ng, SMDA) and three machine-learning-based (DeepDi, RNN, XDA) function start detection tools.
Among the tested tools, IDA achieves the highest \fscore{} (98.44\%) for Chromium x64, while DeepDi closely follows (97\%) but stands out as the fastest by a significant margin.

Working towards explainability, we examine the impact of padding between functions on the detection results.
Our analysis shows that all tested tools, except rev.ng, are susceptible to randomized padding.
The randomized padding significantly diminishes the effectiveness for the RNN, XDA, and Nucleus.
Among the learning-based tools, DeepDi exhibits the least sensitivity and demonstrates overall the fastest performance, while Nucleus is the most adversely affected among non-learning-based tools.

In addition, we improve the recurrent neural network (RNN) proposed by Shin et al. and enhance the XDA tool, increasing the F\textsubscript{1}-score by approximately~10\%.

\end{abstract}

\maketitle

\section{Introduction}

Binary code analysis is a building block of several applications addressing threats for Internet users, such as malware analysis or vulnerability research.
One of the essential first steps in analyzing (compiled) applications is detecting functions in the code (\eg as input to decompilation or for function-level similarity methods).
For example, Haq and Caballero~\cite{haq2021survey} find that 30 out of 61 binary similarity methods operate on a function-level granularity and thus rely on function detection as an initial analysis step.
In this scenario, missing a function or incorrectly detecting a false function start might disrupt code similarity pipelines, potentially preventing them from correctly identifying a malware sample as part of a specific malware family. 
Further, without proper function detection, it is hard to assess the important parts for further analysis (\eg to identify the malicious capabilities a malware sample might exhibit).
Hence, reliably extracting functions from compiled binary software is critical for analyzing binary code.

With millions of hash-unique malware samples emerging every year~\cite{AVInstitute}, automation is key to scalable analysis tooling in various use cases, e.g.,~binary similarity and clustering, malware lineage, actor and tool tracking for threat intelligence, or prioritization for dynamic analysis such as sandboxing~\cite{GoogleTI}.
Thus, reliable, fast, and automated function start detection is key. 

Currently, existing tools often use heuristic or pattern-based methods to identify function starts~\cite{idapro, ghidra, smda_github}. 
These heuristics are compiled by experts based on their experiences when analyzing binaries.
Like all heuristics, these approaches might be incomplete and must be updated regularly.
Thus, recent work suggested machine learning-based approaches~\cite{byteweight,rnn,xda,DeepDi} for function detection to increase performance and accuracy and reduce the need for experts to identify new patterns.
However,  Koo \etAl\cite{Koo_Park_Kim_2021} outline challenges and shortcomings when detecting functions in compiled binary code and revisit previous datasets, metrics, and evaluations.
They show that previous work has suffered from effects such as overfitting (\eg due to missing normalization), in the appropriate definition of true negatives, and imbalance due to significant redundancy in the datasets (shared static library), skewing the evaluation~\cite{Koo_Park_Kim_2021}.
Thus, it is unclear whether machine learning-based approaches meet the expectations and can effectively reduce the need for human experts to develop heuristics.
Additionally, related work often primarily covers functions in Linux/ELF samples and only analyzes fewer Windows/PE functions, if any. 
However, with a significant malware set targeting Windows operating systems~\cite{AVInstitute}, evaluating function detection on PE plays an important role.
While the binary code parts of these two formats can use the same instruction set architecture (ISA) (\eg x64), differences exist in the application binary interfaces (ABI) of Linux and Windows.
Factors that may influence the function start detection efficacy include calling conventions and compiler-specific quirks (\eg padding schemes, inline data, and metadata associated with Windows-specific tools like Microsoft Visual Studio).
Thus, it is unclear if and how the proposed methods can be generalized from ELF samples to other formats.

In this work, we shed light on these challenges by comparing \totalnumtools{} (five heuristics-based and three machine-learning-based) function start detection tools.
We specifically focus on 32-bit and 64-bit Windows PE binaries of benign and malicious code examples.
More specifically, based on previous works~\cite{xda,rnn},  we train two machine-learning-based tools on a commonly used dataset~\cite{byteweight} to find function starts.
We then test the efficiency of these tools and six further tools~\cite{idapro, ghidra, nucleus, DeepDi, smda_dis, revng} on a newly built dataset consisting of samples and ground truth.
We evaluate the tools on a Chromium sample for Windows and the \emph{Conti} ransomware (x86 and x64).
On a high level, our results indicate that all tested tools generalize well across file formats but that the effectiveness of machine-learning-based tools collapses when they face toolchain-specific quirks (\ie different padding schemes). 

\smallskip
\noindent
In summary, we make the following contributions:
\begin{itemize}

\item We introduce FuncPEval, a new x86 and x64 Windows PE dataset that contains malicious and benign software samples spanning \totalnumfunctions{} functions.
Using this dataset, we compare \totalnumtools{} tools, proposed within the past decade and commonly used for function start detection in the field. We show that all used tools can generally identify function starts in regular PE files with high precision and recall.

\item Based on previous work, we train two machine learning-based function start detection tools~\cite{xda,rnn} on a commonly used dataset. For further analysis, we optimize the provided methods, improving XDA’s F\textsubscript{1}-score by approximately 10\%.

\item Finally, we demonstrate that modifying the padding between functions in a given sample impacts the function start detection.
For some tools, the effectiveness of function start detection methods relies heavily on the unmodified padding between functions as emitted by standard compilers. When this padding is altered, the effectiveness for learning-based methods deteriorates, with F\textsubscript{1}-scores down 30 to 70 percent points.
Thus, our results indicate that the machine learning approaches might be susceptible to spurious correlation~\cite{dos_and_donts_machine_learning}.

\end{itemize}

\section{Related Work}
\label{sec:related}

Function detection can be categorized into heuristics- and pattern-based detection (e.g.,~signatures on function prologues and epilogues), static analysis techniques (e.g.,~CFG extraction), and machine learning-based approaches described in more detail in the following subsections.

\begin{table*}
    \caption{Function start detection methods and evaluation papers. The amount of PE functions for Nucleus~\cite{nucleus} is an estimate as the number of samples and the description in the paper indicate that the same dataset as in~\cite{andriesse2016} was used. Compilers refers to the compilers that were used to generate the dataset, i.e., GNU Compiler Collection, Clang, Visual Studio, and Intel C/C++ Compiler.}
    \label{tab:approaches}
    \small
    \setlength{\tabcolsep}{2pt}
    \begin{adjustbox}{max width=\textwidth}
    \begin{tabularx}{\textwidth}{@{\extracolsep{\fill}}p{3.2cm}@{\hskip 2pt}p{0.4cm}@{\hskip 2pt}p{0.4cm}@{\hskip 2pt}p{0.4cm}@{\hskip 2pt}p{0.4cm}@{\hskip 4pt}p{1.2cm}r@{\hskip 0.1cm}r@{\hskip 0.1cm}r@{\hskip 0.1cm}r@{\hskip 0.1cm}c@{}}
        \toprule
        \textbf{Tool/Paper} & \multicolumn{4}{c}{\textbf{Compilers}} & \multicolumn{2}{c}{\textbf{Samples}} & \multicolumn{2}{c}{\textbf{Functions}} & \textbf{ML} \\
        \cmidrule(lr){2-5} \cmidrule(lr){6-7} \cmidrule(lr){8-9}
         & \textbf{GCC} & \textbf{Clang} & \textbf{VS} & \textbf{ICC} & \raggedleft\textbf{ELF} & \textbf{PE} & \raggedleft\textbf{ELF} & \textbf{PE} & \\
        \midrule
        FuncPEval (our dataset) &  & \cmark & \cmark &  & \raggedleft 0 & \raggedleft 4 & \raggedleft 0 & \raggedleft 1,092,820 & \xmark \\
        \midrule
        FunProbe~\cite{FunProbe} & \cmark & \cmark &  &  & \raggedleft 19,872 & \raggedleft 0 & \raggedleft 3,064,001 & \raggedleft 0 & \xmark \\
        DeepDi~\cite{DeepDi} & \cmark &  & \cmark &  & \raggedleft 1,440 & \raggedleft 688 & \raggedleft n/a & \raggedleft n/a & \cmark \\
        Koo et al.~\cite{Koo_Park_Kim_2021} & \cmark & \cmark &  &  & \raggedleft 152 & \raggedleft 0 & \raggedleft 769,069 & \raggedleft 0 & \xmark \\
        FETCH~\cite{pang2021} & \cmark &  &  &  & \raggedleft 43 & \raggedleft 0 & \raggedleft 1,105,278 & \raggedleft 0 & \xmark \\
        XDA~\cite{xda} & \cmark &  & \cmark & \cmark & \raggedleft 2,593 & \raggedleft 528 & \raggedleft n/a & \raggedleft n/a & \cmark \\
        Jima~\cite{Alves-Foss_Song_2019} & \cmark & \cmark &  & \cmark & \raggedleft 3,790 & \raggedleft 0 & \raggedleft 4,913,753 & \raggedleft 0 & \xmark \\
        LEMNA~\cite{lemna} & \cmark &  &  &  & \raggedleft 2,064 & \raggedleft 0 & \raggedleft n/a & \raggedleft 0 & \cmark \\
        Nucleus~\cite{nucleus} & \cmark & \cmark & \cmark &  & \raggedleft 324 & \raggedleft 152 & \raggedleft n/a & \raggedleft est. 378,965 & \xmark \\
        REV.NG~\cite{revng} & \cmark & \cmark &  &  & \raggedleft 1,890 & \raggedleft 0 & \raggedleft n/a & \raggedleft 0 & \xmark \\
        Andriesse et al.~\cite{andriesse2016} & \cmark & \cmark & \cmark &  & \raggedleft 829 & \raggedleft 152 & \raggedleft 1,525,024 & \raggedleft 378,965 & \xmark \\
        Shin et al.~\cite{rnn} & \cmark &  & \cmark & \cmark & \raggedleft 2,064 & \raggedleft 136 & \raggedleft 598,359 & \raggedleft 187,836 & \cmark \\
        BAP/ByteWeight~\cite{byteweight} & \cmark & \cmark &  &  & \raggedleft 2,064 & \raggedleft 136 & \raggedleft 598,359 & \raggedleft 187,836 & \cmark \\
        Rosenblum et al.~\cite{rosenblum} & \cmark &  & \cmark & \cmark & \raggedleft 728 & \raggedleft 443 & \raggedleft 283,626 & \raggedleft 100,427 & \cmark \\
        \bottomrule
    \end{tabularx}
    \end{adjustbox}
\end{table*}

\subsection{Static code analysis and pattern-based approaches}
IDA Pro~\cite{idapro} uses proprietary patterns to identify function starts.
Similarly, Ghidra~\cite{ghidra} uses a combination of signatures\footnote{Ghidra's patterns are available from \url{https://github.com/NationalSecurityAgency/ghidra/blob/b9496de7f573e6a73888abfb51c243723785dbdb/Ghidra/Processors/x86/data/patterns/x86win_patterns.xml}} and static analysis techniques. 
The signatures typically cover machine code instruction sequences in the form of byte patterns that precede a function, such as padding or an epilogue, or that are typically observed at the start of a function, such as a prologue or allocation routines.
In September 2022, Ghidra introduced the Random Forest Function Finder Plugin, a machine learning-based approach that is trained on previously recognized functions of the currently analyzed binary and attempts to find similar function starts.
This differs from the following machine learning approaches in that it is applied on a per-sample scope and does not attempt to globally model function starts.
Andriesse et al.~\cite{andriesse2016} evaluate existing disassemblers and their function boundary recovery.
They conclude that false negative rates of function starts typically reach 20\% or more, indicating significant shortcomings when applied in practice.
A follow-up work by Andriesse et al.~\cite{nucleus} proposes Nucleus, a compiler-agnostic function detection tool that constructs an inter-procedural control flow graph.
Similarly, rev.ng by Di Federico et al.~\cite{revng} proposes a set of analyses to extract function starts based on QEMU's lifter and LLVM's intermediate representation, thus operating without ISA-specific heuristics.

Alves-Foss and Song~\cite{Alves-Foss_Song_2019} propose detecting function boundaries using control flow analysis, jump and call targets, exception metadata, and detection of terminal and missing functions.
Their approach is implemented in the Jima tool, which supports Linux ELF samples only and is currently only available in compiled form.
FETCH by Pang et al.~\cite{pang2021} leverages call frames, i.e., frame description entries in the exception handling information as mandated by the x64/amd64 System V Application Binary Interface.
Such call frame information is typically added by the compiler at build time.
While evaluating against x64 ELF samples only, the authors note that x64 PE and ARM will likely exhibit similar metadata.
However, such metadata is not always present, especially in malware.

SMDA by Plohmann~\cite{smda_dis} combines recursive disassembly and heuristics for function entry point discovery and later performs a gap analysis to find missed functions.

In 2023, FunProbe by Kim et al.~\cite{FunProbe} proposes a probabilistic model using a Bayesian Network over causal relationships between heuristically identified function entry point candidates.
First, an inter-procedural CFG is recovered, and up to 16 function identification hints are collected based on data-driven properties, e.g.,~FDE, and code-driven properties, e.g.,~call targets.
Then, a Bayesian Network is built, followed by belief propagation.
As a result, each byte yields an inferred probability of being a function start when exceeding a given threshold.
FunProbe currently only supports ELF files and has been evaluated on a total of 19,872 ELF samples covering x86, x64, ARM, and MIPS architectures.

\subsection{Machine learning based approaches}
Approaches that leverage machine learning have initially been proposed by Rosenblum et al.~\cite{rosenblum}.
They model function start detection as a classification problem using Conditional Random Fields.

In 2014, Bao et al.~introduced a function boundary detection approach called ByteWeight~\cite{byteweight}.
It uses a weighted prefix tree to learn signatures that can be used to detect function starts.
In the case of ByteWeight, each branch represents a sequence of bytes or instructions.
The depth of the tree determines the length of the sequence.
The weighted prefix tree adds a weight to each node in the tree, representing the probability that the branch starts a function.
The authors build the weighted prefix tree by training a non-weighted prefix tree with all possible byte or instruction combinations using ground truth data.
Therefore, ByteWeight suffers from the same problems as traditional signature-based approaches, e.g.,~depending on compiler versions.
Nevertheless, ByteWeight can automatically generate signatures when ground truth data is available.
For each architecture and compiler, ByteWeight has to generate new signatures and, therefore, also requires new ground truth data.
While ByteWeight aims to detect function starts, Bao et al.~present further analysis techniques that can be applied after the function start detection that lift the approach to detect all instructions belonging to the function.

In 2015, Shin et al.~\cite{rnn} introduced function boundary detection using a bidirectional recurrent neural network (RNN).
The RNN uses a sequence of bytes as input and decides for each byte if it marks the start of a function.
Similar to ByteWeight, the weights of the RNN are trained using ground truth data.
The trained weights form the model used to detect function starts.
The RNN can also detect the boundaries of a function by using two models. 
One model detects function starts, and the other model detects function ends.
The authors do not provide an implementation of their approach.
However, a reimplementation was provided by Guo et al.~\cite{lemna} as part of their work on the explainability of machine learning-based methods.

In 2021, Pei et al.~\cite{xda} propose XDA, which relies on transfer learning of machine code disassembly and also recovers function boundaries.
They are motivated by masked language modeling to infer dependencies between specific bytes in machine code.
While the paper evaluates on x86 and x86-64 samples, targeting both ELF and PE, a fine-tuned model of XDA has only been published for x86-64.
XDA models function boundary detection as a multi-class classification problem where a specific byte can either form the start of a function, the end of a function, or neither of both.

DeepDi, a system published by Yu et al.~\cite{DeepDi} in 2022, combines instruction-level sequences with a graph convolutional network to achieve disassembly.
First, given a byte string as input, all possible instructions are decoded using a 15-byte sliding window over the input stream, yielding the superset of instructions.
Then, an instruction flow graph is constructed that captures the most likely true instructions and their relations.
The system also contains heuristics and a classifier for function start detection based on the resulting disassembly.
In contrast to previous work, DeepDi is the first learning-based approach that operates on the instruction level instead of the byte level.

\subsection{Limited focus on PE files in existing work}

\Cref{tab:approaches} summarizes the related work, including details such as the number of samples and functions used for evaluation, where available.
These figures are presented separately for ELF and PE binaries.
The table reveals that only 6 out of 13 studies evaluated PE samples.
Comparing the number of samples and functions between ELF and PE in these six cases highlights a significant underrepresentation of PE binaries in the existing literature.
Given the importance of function detection in PE samples, particularly in the context of malware analysis~\cite{AVInstitute}, it becomes evident that a new evaluation focusing on PE binaries with a larger set of functions is necessary.

\section{Background on Function Detection}
\label{sec:func}

\begin{figure}
	\begin{subfigure}[b]{\linewidth}
		\centering 
		\includegraphics[width=\textwidth]{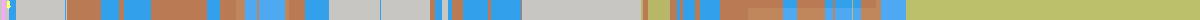}
		\caption{IDA Pro. blue: code in functions, brown: instructions outside of functions, grey: data, amber: unexplored.}
		\label{fig:urausy_comparison_a}
	\end{subfigure}%
	\qquad
	\begin{subfigure}[b]{\linewidth}
		\centering 
		\includegraphics[width=\linewidth]{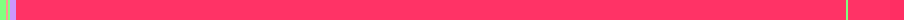}
		\caption{Ghidra. red: undefined data, green: data, purple: code in functions.}
		\label{fig:urausy_comparison_b}
	\end{subfigure}%
	\caption{Comparison of code, data, and unexplored areas in memory-mapped views of a Urausy malware sample.}
	\label{fig:urausy_comparison}
\Description[]{}
\end{figure}

Function detection in compiled code is not straightforward.
Different approaches to function detection are likely to yield varying sets of recognized functions stemming from the diverse methodologies employed.
The following example highlights the significant differences in the detection of functions in a malware analysis context.
\autoref{fig:urausy_comparison} shows two memory mapped representations of a Urausy malware sample\footnote{SHA256 hash value of \ttfamily\footnotesize\seqsplit{8f4296a0990ec245997bd2bb75edb512aae4e544b7d0c36e945bf19241fda426}}, produced by IDA Pro~\cite{idapro} and Ghidra~\cite{ghidra}, two popular reverse engineering tools.
The colors represent the types of data or code when mapped in memory.
The tools differ significantly regarding the detected functions.
\autoref{fig:urausy_comparison_a} shows the analysis results of IDA Pro where
blue-colored areas represent detected functions, brown-colored areas represent instructions that do not belong to functions, gray-colored areas represent data, and amber-colored areas represent unexplored areas that could not be further specified by IDA.
In contrast, \autoref{fig:urausy_comparison_b} shows the analysis results of Ghidra where purple-colored areas represent detected functions, green-colored areas represent data, and red-colored areas represent undefined data.
The figures show that the tools differ significantly in the detected functions in this sample.
Manual analysis of this sample reveals that the instructions that IDA classified as not belonging to functions actually belong to functions.
Similarly, Ghidra misclassified such code as undefined data.
In practice, an analyst would need to inspect the code and manually define functions that were missed during function start detection, a labor-intensive task.
The example shows that the function detection problem is far from being solved and that an evaluation of the existing tools is appropriate and needed.

Following most related work, we refer to function detection as finding bytes in compiled binary code that belong to the same function in the source code and for which no symbol information is given.
There are other terms that refer to the same problem, e.g.~function boundary detection, function boundary identification, function identification, and function recognition~\cite{nucleus,byteweight,Koo_Park_Kim_2021}.
We will use the term function \emph{detection} to avoid confusion with library function recognition~\cite{binshape,library_function_identification}.
This different problem may occasionally also be referred to as function identification and describes recovering the semantic meaning of a function given its binary code, which is out of the scope of this paper.

Furthermore, we distinguish between i)~function start detection, which only considers the \emph{start} of functions, and ii)~function boundary detection, which typically detects function \emph{start} and \emph{end} addresses, and iii)~code ranges covering intervals of function code, most relevant for non-continuous functions.

Unless stated otherwise, we focus on function \emph{start} detection for the remainder of this work because it applies to all related work and thus allows us to include the most tools in our evaluation.
Furthermore, it can be modeled as a binary classification problem, enabling the use of well-known and considered evaluation metrics.

\begin{figure}
\includegraphics[width=\linewidth]{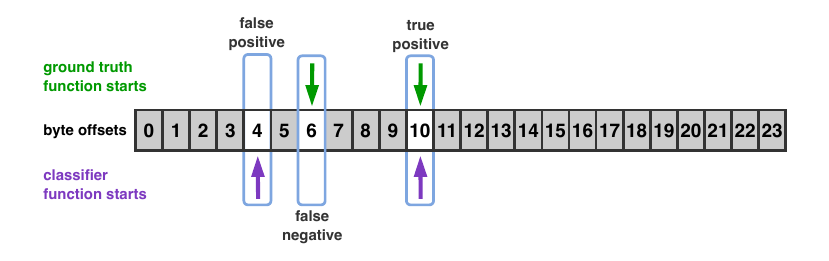}
\caption{Schematic of function start detection} \label{fig:funcstart}
\Description[]{}
\end{figure}

To evaluate function start classifiers, we use the \emph{precision}, \emph{recall}, and \emph{F\textsubscript{1}-score} as metrics.
To illustrate, \autoref{fig:funcstart} depicts a schematic, fictional binary of size 24 bytes, alongside a hypothetical classification of function starts.
Here, the ground truth marks the bytes at offsets \texttt{6} and \texttt{10} as valid function starts (\eg obtained via debugging symbols).
In contrast, the hypothetical classifier considers the bytes at offsets \texttt{4} and \texttt{10} to be function starts.
True positives (TP) refer to bytes correctly identified as function starts, matching the ground truth, \eg the byte at offset \texttt{10} in \autoref{fig:funcstart}.
False positives (FP) are bytes incorrectly classified as function starts, as they do not correspond to function starts in the ground truth (\eg byte \texttt{4} in \autoref{fig:funcstart}).
False negatives (FN) are bytes that are not classified as function starts but are function starts according to the ground truth (\eg all grey-shaded bytes in \autoref{fig:funcstart}).

Note that in \autoref{fig:funcstart}, only two out of 24 bytes represent function starts.
This highlights a common characteristic: the number of function starts rarely exceeds a small fraction of the total number of bytes in the file.
Since a binary consists of much more than function starts, and each function start is represented by only a single byte, this imbalance is likely prevalent in most datasets.
Such imbalance must be accounted for during training (\eg by initializing the biases accordingly) and evaluation (\eg by avoiding accuracy as a metric).

As previously described, most bytes in our context will not represent a function start.
Given this imbalance, we avoid \textit{accuracy} as a metric because it factors in true negatives in the computation.
The example in \autoref{fig:funcstart} would yield an accuracy of 92\%, and an \fscore{} of 50\%.
Similarly, while practically useless, a naive classifier that predicts \emph{every} byte as a non-function start would still result in a (comparatively) high number of true negatives and, consequently, high accuracy.
Arp et al.~\cite{dos_and_donts_machine_learning} refer to such a pitfall as inappropriate performance measures.

\section{Empirical Validation}
\label{sec:reproduction}

Our work aims to assess all function start detection tools introduced within the past decade that can operate on PE files.
To achieve this, we describe the BAP/ByteWeight dataset as it is typically used to train models, and develop models for two learning-based approaches.
We reproduce the work of Shin et al.~\cite{rnn}, as their original implementation is not publicly available, and existing reimplementations by other researchers~\cite{xda, lemna} do not include trained models.
Therefore, we provide a stable, well-documented implementation of Shin's RNN-based classifier and publish the x86 and x64 Windows PE training data along with the trained models.

Additionally, we discovered a discrepancy in the implementation of XDA~\cite{xda}, leading to the reproduced results deviating from those reported in the original paper.
To rectify this inconsistency, we introduce a novel encoding for the training data, leading to results that more closely align with those presented in the paper.

\subsection{BAP/ByteWeight Dataset}
\label{sec:bwdataset}
To reproduce the original implementations of Shin’s RNN and XDA, we use the same dataset for training and evaluation, as used in the original studies.
The dataset was initially compiled as part of BAP/ByteWeight~\cite{byteweight}.
We only use a subset of the dataset, \ie PE binaries targeting Microsoft Windows and their corresponding ground truth files containing the function starts as virtual addresses (VA).
In the following, we refer to it as BAP dataset.
This dataset spans a total of 136 hash-unique PE samples, built using Microsoft Visual Studio versions 2010 to 2013.
It consists of 68 x86 and 68 x64 PE samples compiled from 17 programs (7z, vim, various PuTTy tools, hidapi, libsodium, sfxsetup, smtpsend), using four optimization levels (Od, O1, Ox, O2).
In one case (filename \filename{msvs\_whatever\_32\_Od\_SfxSetup}), the ground truth in the dataset did not match the provided sample as it contains a function VA where no section is mapped in memory.
This could potentially be due to a misunderstanding or an error made by the original authors.
To avoid label inaccuracy~\cite{dos_and_donts_machine_learning}, we exclude this specific x86 sample from our dataset, resulting in a total of 135 samples (67 x86 and 68 x64).

Koo et al.~\cite{Koo_Park_Kim_2021} highlight a pitfall of the BAP/ByteWeight dataset, particularly in the ELF subset, that we do not use.
When normalizing instructions by blinding immediates and call and jump targets, only 17.6K (12.1\%) out of the whole 146K functions of the ELF subset form unique normalized functions~\cite{Koo_Park_Kim_2021}.
In other words, without normalization, overfitting becomes likely, especially when a significant number of normalized functions is present in both the training and the validation set.
We decided against normalization in the training part of our pipeline to keep our results comparable with prior work.
However, we consider uniqueness and normalization in our new FuncPEval dataset introduced in \Cref{sec:replication}.

\subsection{Implementation of the RNN-based Classifier}
\label{sec:implrnn}
\begin{figure}
	\begin{subfigure}[b]{\linewidth}
		\centering 
		\includegraphics[width=\linewidth]{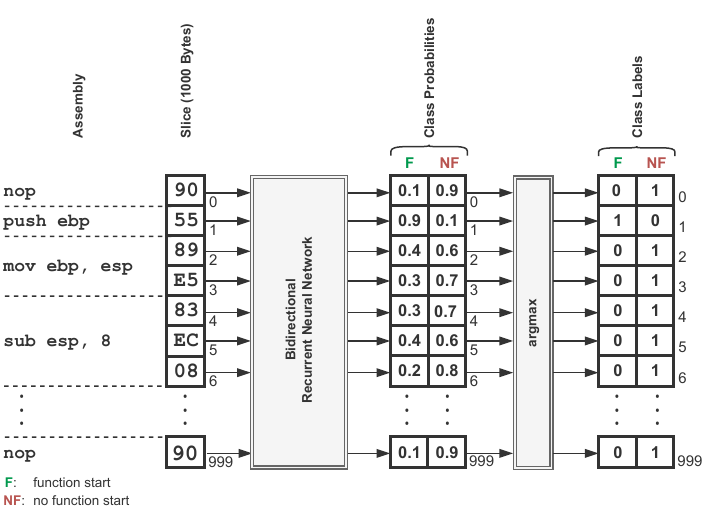}
		\caption{LEMNA-based RNN pipeline with two output neurons}
		\label{fig:orign_rnn_arch}
	\end{subfigure}%
	\qquad
	\begin{subfigure}[b]{\linewidth}
		\centering 
		\includegraphics[width=\linewidth]{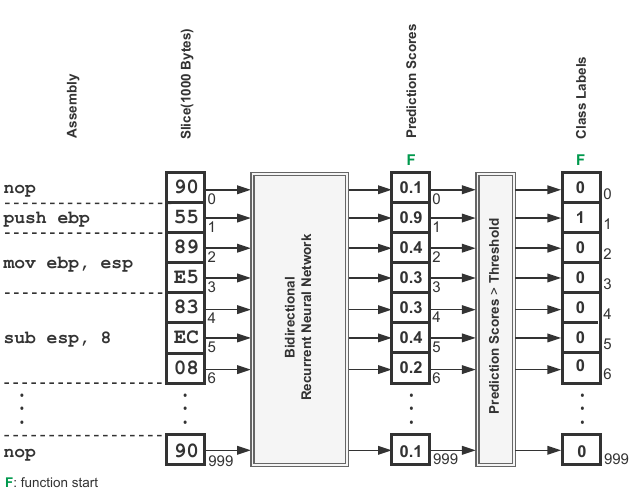}
		\caption{Our modified RNN pipeline with one output neuron}
		\label{fig:mod_rnn_arch}
	\end{subfigure}%
	\caption{Comparison of the original and the modified RNN pipelines} 
	\label{fig:arch_comparison}
\Description[]{}
\end{figure}

Currently, no public implementation of Shin’s RNN includes trained models for the classification of PE files.
To incorporate the RNN into our evaluation in \Cref{sec:replication}, we reproduce the original implementation and train models for x86 and x64 PE.
Based on previous work by Shin et al.~\cite{rnn} and using the LEMNA reimplementation by Guo et al.~\cite{lemna} as basis, we implement the RNN-based classifier in Python using Keras~2.8.0 and TensorFlow~2.8.0 as backend.
Considering the function start detection as a binary classification problem, we also implement a second variant with a slightly modified pipeline, namely one output neuron. 
In contrast, the LEMNA reimplementation uses two output neurons (\autoref{fig:orign_rnn_arch}).
Our modification allows us to adapt the trigger threshold and thereby improve the classification results as shown in \Cref{tab:exp1results}.
The LEMNA-based RNN pipeline and our modified pipeline are shown in \autoref{fig:arch_comparison}.

Shin's RNN architecture specifies that a sample must be divided into 1000-byte slices.
If the last slice is shorter than 1000 bytes, the slice will be padded with zeros.
Subsequently, the slice is fed into the RNN,
where i)~the LEMNA-based RNN outputs two class probabilities for each byte, and ii)~our modified RNN yields a prediction score for each byte.
Thus, we can predict for each input byte if it is the start of a function in the binary.
Finally, a mathematical function assigns a class label to the byte. 
Guo et al.~\cite{lemna} use $argmax$, which determines the class label based on the maximum of the two class probabilities.
Our variant uses a heuristic thresholding function:
\begin{equation}
	C_{{\text{Label}}}(P_{{\text{Score}}})={\begin{cases}0&{\text{if }}P_{{\text{Score}}}\leq t\\1&{\text{if }}P_{{\text{Score}}}> t\end{cases}}
\end{equation}
If the prediction score $P_{Score}$ exceeds the threshold $t$, the class label $C_{Label}$ for the byte will be set to $1$ (\ie function start), otherwise $0$ (\ie not a function start).
To determine the optimal threshold, the training dataset was evaluated with threshold values from $0$ to $1$ at intervals of $0.01$. 
Subsequently, the threshold ($0.38$) with the best $F_1$-score was selected and used for the remaining experiments.

Note that when considering function starts on a per-byte level, every dataset containing real-world binaries is imbalanced as shown in \Cref{sec:func}, 
because bytes that do \textit{not} represent a function start appear more often than those that \textit{do} represent a function start.
We randomly initialize the weights of the RNN and take the imbalance into account by changing the bias of the output layer.
The initial bias $b_0$ is computed as follows:

\begin{equation} 
	b_0 = log_{e}\left(\frac{pos}{neg}\right)
\end{equation}

where $pos$ is the number of bytes representing the start of a function and $neg$ is the number of bytes that are \emph{not} the start of a function.
Shin et al.~\cite{rnn} use the adaptive learning rate optimizer RMSprop~\cite{hinton} for training, while we use Adam~\cite{adam}, the same used in LEMNA.

\subsubsection{Training and Validation}
\label{sec:train}

We train and validate separately for x86 and x64 PE samples.
Given the BAP dataset of 67 x86 PE samples, we perform 10-fold cross-validation as follows:
Split the dataset into ten (nearly) equally-sized disjoint subsets where each subset spans ca.~10\% of the samples.

In each fold, executable sections from 9 of the 10 subsets (ca.~90\% of the samples) are used for training, and executable sections from the remaining subset (ca. 10\% of the samples) are used for validation.
Shin et al.~use a batch size of 32, while Guo et al.~\cite{lemna} used 100.
We use a batch size of 1,000 for the given dataset to increase training speed.
While Shin et al.~trained for two hours, we followed LEMNA and trained the RNN for 150 epochs for each fold (taking approximately 55 minutes per fold).
We run our experiments on a server with two 2.20 GHz Intel Xeon Silver 4114 CPUs à 20 threads and 128 GB of RAM.

\begin{table*}
    \caption{Results from 10-fold cross-validation of our RNN models compared to previous work by Shin et al., using the BAP PE dataset. Shin et al.~report higher F\textsubscript{1}-scores, but their implementation and trained models are unavailable for inspection.}
    \label{tab:exp1results}
    \begin{tabularx}{\textwidth}{lXcccXccc}
        \toprule
                                  & &           & \textbf{PE x86}         &                      &  &                & \textbf{PE x64}    &  \\ 
        \textbf{Method}                               & & Precision & Recall         & F\textsubscript{1}-score &  & Precision  & Recall       & F\textsubscript{1}-score  \\
        \midrule
        Shin et al.~\cite{rnn}              & & 99.01\%   & 98.46\%        & 98.74\%              &  & 99.52\%        & 99.09\%      & 99.31\%  \\
        \midrule
        two output neurons RNN                   & & 97.41\%   & 92.42\%        & 94.83\%              &  & 98.66\%        & 96.43\%      & 97.53\%  \\
        one output neuron RNN                   & & 96.96\%   & 95.65\%        & 96.29\%              &  & 98.62\%        & 98.29\%      & 98.45\%  \\
        \bottomrule
    \end{tabularx}
\end{table*}

\subsubsection{Evaluation}
\label{sec:evalexp1}

Table~\ref{tab:exp1results} shows the precision, recall and F\textsubscript{1}-score, on average.
Although they do not exactly match the values from the previous work by Shin et al., we consider the results similar, indicating that our implementation provides a suitable RNN-based function start classifier.
Note that our modified variant, referred to as one output neuron RNN, achieves higher F\textsubscript{1}-score values for both x86 and x64, compared to the two output neurons RNN.
As a result, we use the one output neuron RNN in subsequent experiments.

In the original paper, Shin et al.~yield higher F\textsubscript{1}-scores compared to our models.
However, we lack a precise explanation, as their implementation and trained models are inaccessible.
This may be an artifact of different folds or implementation details.
For the experiments in \Cref{sec:replication}, we train models for PE x86 and x64 using the whole BAP dataset.
We publish our documented implementation, training data, and models.

\subsection{Improving XDA}
\label{xda_reproduction}
In preparation for the tool comparison in \Cref{sec:replication}, we noticed discrepancies in the encoding of function starts and ends in the published XDA tooling.
Consequently, we reproduce the results to verify their correctness.
We utilized the model and code published by the authors for 64-bit PE files and applied it to the x64 part of the BAP dataset (BAP-64).
In the original evaluation, 90\% of the BAP dataset was used for evaluation because 10\% was used to train XDA.
Since we do not know which samples were used for training and which for evaluation, we use the entire dataset for evaluation.
This should only positively impact the results for XDA, as 10\% of the dataset was already seen during training.
First, the function starts and ends are predicted using the provided model.
Subsequently, we combine these starts and ends into function boundary pairs using the published algorithm\footnote{\label{foot:XDA_pairing_1}\url{https://github.com/CUMLSec/XDA/blob/main/scripts/play/eval_pair_bound.py\#L6}, Commit: c3cce2f}.
Following this, we utilized the F\textsubscript{1}-score calculation provided by the authors\footnote{\url{https://github.com/CUMLSec/XDA/blob/main/scripts/play/eval_pair_bound.py\#L39}, Commit: c3cce2f} to closely align with the original evaluation.
The computed F\textsubscript{1}-score is shown in \Cref{tab:xda_experiment} in the column named \emph{our experiment}, and the F\textsubscript{1}-score from the original work in the column named \emph{reported}.

We also include Nucleus, Ghidra, and IDA in the evaluation to provide a better comparison with the original evaluation~\cite{xda}.
We chose not to include the bi-RNN, as the XDA authors did not provide trained models for its implementation, making a fair comparison impossible.

\begin{table}
\centering
\small
\caption{Improving XDA on the BAP-64 PE dataset. The column \emph{our experiment} shows the results in our reproduction, the column \emph{reported} shows the results as presented in~\cite{xda}, and \emph{adapted GT} shows the results evaluated against the adapted ground truth, which was potentially used in~\cite{xda}}
\label{tab:xda_experiment}
\begin{tabular}{lccc}
    \toprule
    & \multicolumn{3}{c}{\textbf{F\textsubscript{1}-score (PE x64)}} \\
    \textbf{Tool} & \makecell[c]{our \\ experiment} & reported~\cite{xda} & adapted GT \\
    \midrule
    IDA           & 91.13\%  & 90.5\%  & 78.66\%   \\
    Ghidra        & 78.69\%  & 80.6\%  & 71.22\%   \\
    Nucleus       & 79.80\%  & 70\%    & 67.55\%   \\
    XDA reproduced & 82.68\%  & 99.4\%  & 97.81\%   \\
    XDA new encoding & 93.66\%  & -       & -         \\
    \bottomrule                      
\end{tabular}
\end{table}

The results in Table~\ref{tab:xda_experiment} indicate that the outcomes of our evaluation for IDA and Ghidra are similar to those in the original evaluation.
The slight deviation could be attributed to the use of a slightly different dataset (100\% of BAP-64 in our evaluation instead of 90\%).
However, there is a significant discrepancy between our evaluation and the original evaluation in the results for XDA (\verb|~|17 percent points in \fscore) and Nucleus (\verb|~|10 percent points in \fscore).
To investigate the cause of this discrepancy, we conducted a detailed analysis of XDA’s detection mechanism.

\subsubsection{XDA Label Encoding}
\label{sub:xda_label_encoding}
\begin{figure}
\includegraphics[width=\linewidth,trim={0 0 2.3cm 0},clip]{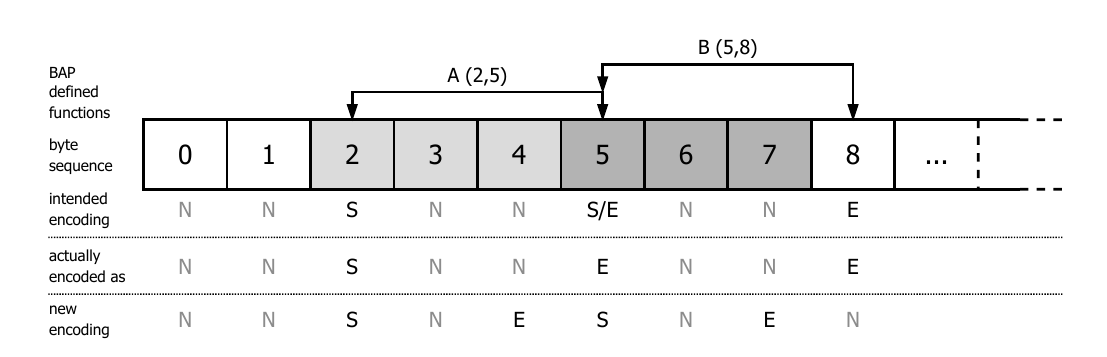}
\caption{Labeling of Functions in XDA. For two adjacent functions, XDA would need to assign two labels for one byte. That is impossible, and a new encoding is required.} \label{fig:xda_labeling}
\Description[]{}
\end{figure}

Upon closer examination of the functions detected by XDA, we noticed that XDA fails to correctly identify functions when they are immediately adjacent to each other, \ie when one function ends and another begins directly afterward without any bytes in between.
This issue stems from the labeling of functions in the BAP dataset.
In the BAP dataset, each function is labeled with a start and an end.
The start is inclusive, marking the first byte of the function, while the end is exclusive, indicating the first byte that does not belong to the function.
When one function directly follows another without intervening bytes, the end of the first function is the same as the start of the second function in the BAP notation.
XDA assigns exactly one label per byte, \emph{S} for a function start, \emph{E} for a function end, and \emph{N} for neither.
In the scenario of two adjacent functions, XDA either correctly identifies the end of the first function \emph{or} the start of the second function.
Therefore, XDA cannot identify both functions in such cases.
Figure~\ref{fig:xda_labeling} illustrates this issue.
Function B (5,8) directly follows function A (2,5).
XDA would need to assign both, the label \emph{S} and the label \emph{E}, to byte 5 to identify both functions correctly.
If XDA classifies byte 5 as the end of a function, there are two function ends, i.e., byte 5 and byte 8, without a function start in between, and XDAs pairing code yields the function boundary pair (2,8), effectively representing a single function starting at byte 2 and ending at byte 8.
This would result in one false positive and two false negatives in the evaluation.
Note that we do not consider this a shortcoming of the XDA classifier but rather a labeling issue.

In the published artifacts, this issue also affects the training data for XDA.
Each byte is assigned exactly one label, and the label \emph{E} is assigned in case of a label conflict.
Consequently, XDA never encounters a function start that immediately follows the end of another function during training.
Additionally, the published code\footnote{\url{https://github.com/CUMLSec/XDA/blob/5315918317eda39bf5de8ca56935baabfc30aa7e/scripts/play/eval_pair_bound.py\#L144}, Commit: c3cce2f} suggests that the ground truth data for the evaluation was encoded similarly, causing the ground truth in XDA to deviate from the BAP ground truth.
We attempted to reconstruct the ground truth as used in the original work by i)~dividing the set of functions in the BAP ground truth into pairs of function starts and ends, ii)~extracting all function starts in the set of function starts that also appeared in the set of function ends, and iii)~forming new function boundary pairs by using the published algorithm\footnote{\label{foot:XDA_pairing_2}\url{https://github.com/CUMLSec/XDA/blob/main/scripts/play/eval_pair_bound.py\#L6}, Commit: c3cce2f}.

With the adapted ground truth, the evaluation results for XDA and Nucleus are much closer to those in the original evaluation, as shown in \Cref{tab:xda_experiment} in the rightmost column named \textit{adapted GT}.
The minor discrepancies could be attributed to the slightly different datasets used (100\% of BAP-64 in our evaluation instead of 90\%).
However, the results for IDA and Ghidra deviate significantly from the original evaluation compared to the BAP ground truth.
Although speculative, one possible explanation is that different tools might have been evaluated differently, using the adapted ground truth for XDA and Nucleus while using the original BAP ground truth for IDA and Ghidra.

\subsubsection{Retraining XDA}
\label{sub:retraining_xda}
We aim to address the labeling inaccuracy of the training data encoding by slightly modifying the training data and retraining XDA.
In the new encoding, we treat the end of a function as inclusive, just like the start, marking it as the last byte that still belongs to the function (\emph{new encoding} in \Cref{fig:xda_labeling}).
During evaluation, we account for this new encoding method by increasing the value of each function end by one.
This adjustment is also applied when evaluating Ghidra and Nucleus, as both also consider the end of the function to be inclusive.
With the help of the new encoding, only labels for one-byte-sized functions would result in a label conflict.
In these cases, it is impossible to assign a single correct label, as the first and only byte of the function represents both the start and the end.
To address this, a new label would need to be introduced to mark a function's start \emph{and} end.
Since single-byte-sized functions are extremely uncommon, only occurring in about 0.1\% of the functions in the BAP-64 dataset, we decided not to implement this change.

We use a randomly selected 10\% of the samples from the BAP-64 dataset to retrain (fine-tuning part only) XDA and evaluate the entire BAP-64 dataset.
This should not negatively impact the results of XDA, as 10\% of the evaluation dataset was already seen during training.

This newly trained version of XDA achieves significantly better results than the original version, as shown in the last row of Table~\ref{tab:xda_experiment}, with an increase from 82.68\% to 93.66\% in the F\textsubscript{1}-score.
The F\textsubscript{1}-score of the newly trained version is still more than 5 percent points lower than the value reported in the original paper~\cite{xda}.
However, in our evaluation, the retrained version of XDA again emerges as the tool with the best results compared to IDA, Ghidra, and Nucleus.
This supports our assumption that the XDA approach fundamentally works well, although it does not quite meet the claims of the original paper.
To investigate this further, the evaluation could be repeated on other datasets, i.e., SPEC2006, SPEC2017, and the x86 versions.
However, the pre-trained models used in the original evaluation would need to be made available for a fair comparison.
Our experience underlines the importance of reproducible artifact and dataset publication in our research community to better explain such differences.

\begin{lesson}
This section shows how Shin's RNN can benefit from using a one-output-neuron pipeline yielding F\textsubscript{1}-scores of 96.29\% for PE x86, and 98.45\% for PE x64. 
Similarly, XDA significantly benefits from a different labeling scheme, improving its F\textsubscript{1}-score by nearly 11 percent points for PE x64.
Adjusting the labeling to account for adjacent functions underscores the critical role of thoroughly understanding and modeling the problem domain.
\end{lesson}

\section{Comparing Function Start Detection Tools}
\label{sec:replication}

As demonstrated in \Cref{sec:related}, prior research primarily focused on evaluating function start detection tools using ELF binaries.
In contrast, our objective is to evaluate function start detection tools developed over the past decade exclusively on PE binaries.
We introduce and release a new Windows PE dataset called FuncPEval to facilitate this.
This new PE evaluation dataset spans 549k functions (233k normalized functions) for x86 and 543k functions (316k normalized functions) for x64, more than twice compared to previously available PE datasets.
As a result, our dataset allows comparing tool performance using F\textsubscript{1}-score and execution speed.
Additionally, we investigate the impact of padding bytes between functions on function start detection.

\subsection{FuncPEval}
\label{sec:evaldataset}

\begin{table*}
    \centering
    \caption{Properties of the training and the evaluation datasets, showing the number of functions and prologues. 
    Normalization includes blinding immediates as well as call and jump targets. 
    }
    \label{tab:evaldataset}
    \begin{tabular}{llrp{0.1pt}rrrp{0.1pt}rrrp{0.1pt}r}
        \toprule
        & & & & \multicolumn{3}{c}{\textbf{Functions}} & & \multicolumn{3}{c}{\textbf{Prologues (\texttt{.pdata})}} & & \multicolumn{1}{c}{\textbf{Padding}} \\
        \textbf{Dataset} & \textbf{PE sample(s)} & \textbf{Arch} & & RVAs & unique & norm. & & present & unique & norm. & & instances \\
        \midrule
        BAP PE        & 68 samples & x64 & &  94,548 &  65,733 (70\%) &  18,169 (19\%) & &  74,057 &  10,979 & 1,775 &  &  84,069     \\
        \midrule
        FuncPEval   & Chromium v109 & x86 & & 548,534 & 541,707 (99\%) & 232,781 (42\%) & &       ~ &       &      & & 377,769 \\
        FuncPEval   & Chromium v109 & x64 & & 542,902 & 536,182 (99\%) & 315,745 (58\%) & & 470,317 &  7,488 & 1,982 & & 390,063 \\
        FuncPEval   & Conti v3    & x86 & &     722 &    721 (99\%) &    524 (73\%) & &       ~ &       &      & &     310 \\
        FuncPEval   & Conti v3    & x64 & &     662 &    659 (99\%) &    450 (68\%) & &     389 &    179 &   122 & &     548 \\
        \bottomrule
    \end{tabular}
\end{table*}

To address the limitation of prior research focusing predominantly on ELF binaries, we introduce a new dataset, \emph{FuncPEval}, which exclusively comprises PE binaries.
The dataset contains PE binaries targeting Microsoft Windows, stripped off their debugging information.
As benign software, the core library \texttt{chrome.dll} of Chromium version 109 is chosen, both as x86 and x64 PE.
These samples were compiled and linked by Google using LLVM clang 14.10.25019, using various per-module optimization levels, including Ox, Os, Oy, O1, Ot, and O2.
For x64, we use Chromium snapshot 1069922\footnote{\url{http://commondatastorage.googleapis.com/chromium-browser-snapshots/index.html?prefix=Win_x64/1069922/}, chrome.dll has a SHA256 hash value of \texttt{55f05fe24ebdf8eb263f75e88c8a71a42fb6240b59340a9abf9671ffe79a4f4a}}, and for x86, we use Chromium snapshot 1069956\footnote{\url{http://commondatastorage.googleapis.com/chromium-browser-snapshots/index.html?prefix=Win/1069956/}, chrome.dll has a SHA256 hash value of \texttt{1ce8b9551709581688a8199a0e0fcb48cfcac7fadf3671622ea8e66fbe39151f}}, both released 10 Nov 2022.
We choose to include Chromium for two main reasons: First, given its wide adoption, relevance in practice, and diverse code base, Chromium is a suitable target for binary code analysis. 
Second, Chromium for Windows has not been used in previous evaluations, rendering it unlikely that existing tools have been particularly optimized for our dataset.

We extract ground truth on the function start addresses from the associated PDB files using Microsoft's DIA API and consider functions that are designated as symbol type \textit{Function} and are listed under the respective compiled modules (\filename{*.obj}). 
For each module, relative virtual function addresses (RVA) and other symbols (\textit{FuncDebugStart, FuncDebugEnd, etc.}) are specified, which can be found in the PE file after linking the respective modules. 
We ignore all other symbol types, e.g.~\textit{Thunk}, because they do not provide any relevant information in the context of the given problem.

In addition to Chromium, we evaluate against the code of the Conti ransomware, a prevalent malicious software made publicly available by a leak in early 2022~\cite{vxunderground}.
We compiled and linked \textit{Conti} version~3 using Visual Studio~2022 (version 14.34.31933) for both x86 and x64 and generated PDB files to obtain the ground truth.
In the following, we consider the crypter (\texttt{cryptor.exe}), which is the component that encrypts files on the victim machine.
To the best of our knowledge, no prior work has evaluated function start detection using malware in combination with reliable ground truth.
While it would be beneficial to include a broader range of samples and malware families, obtaining reliable ground truth is challenging, as it necessitates access to either a compilable version of the source code or corresponding debugging symbols.

To highlight the diversity in our evaluation dataset, we analyzed the binary code as shown in Table~\ref{tab:evaldataset}. 
For example, Chromium x64 exhibits 542,902 distinct RVAs in its chrome.dll that denote the start of a function.
However, these represent 536,182 byte-unique functions, i.e.~some functions are byte-equal multiples.
ByteWeight~\cite{byteweight} proposed normalizing instructions by removing immediates and call/jump targets.
When normalized, the Chromium x64 sample contains 315,745 distinct normalized functions (roughly 58\% of all functions).
This shows that Chromium exhibits enough diversity for our evaluation.
For reference, the overlap between Chromium x64 and BAP-64 is minimal, consisting of only 95 byte-unique functions (out of 601,820), corresponding to 176 normalized functions.

In addition, we analyzed the function prologues for the x64 samples.
Assuming that function start detection tools may consider the bytes at the beginning of a function, which often represent the function prologue, particularly important, their diversity in the dataset becomes particularly noteworthy.
Hence, the second-rightmost columns in Table~\ref{tab:evaldataset} describe the diversity of prologues.
Information about the prologue is derived from the \texttt{.pdata} section in x64 PE files, and only available for functions that use exception handling
\footnote{See \url{https://learn.microsoft.com/en-us/windows/win32/debug/pe-format\#the-pdata-section}}.
For Chromium x64, a non-zero-sized prologue was present in 470,317 functions, representing 7,488 unique prologue byte sequences.
When normalizing the prologue instructions, 1,982 distinct sequences remain.
While this number is significantly lower than the number of functions that exhibit a prologue, it is expected, as the diversity of prologues is certainly limited in general.
Nevertheless, the number of distinct normalized prologues provides a measure of the diversity of function prologues.
Since the tools might also consider the bytes before the function starts in their detection, we have included the number of functions with at least one padding byte before the function start in the rightmost column of Table~\ref{tab:evaldataset}.

For comparison, the BAP x64 PE dataset contains 65,733 byte-unique functions, i.e., unique function byte sequences.
With normalization, only 18,169 normalized functions (ca.~28\%) remain.
Similarly, out of 10,979 byte-unique prologues based on \texttt{.pdata} section, 1,775 normalized prologues (ca.~16\%) remain.
Both the absolute numbers and the relative numbers, \ie the number of normalized functions over the total number of functions, demonstrate fewer duplicates compared to the BAP dataset, indicating that the FuncPEval dataset contains more diversity, which is a desirable property for an evaluation dataset.

\subsection{Evaluation and Tool Comparison}
\label{sec:exp2}

To assess the performance of function start detection tools on PE files, we utilize our newly introduced comprehensive FuncPEval dataset to evaluate eight tools, measuring their performance in terms of speed and \fscore{}.
The research question is as follows:
How do function detection tools perform when predicting function starts in PE programs with diverse code that exhibit different build toolchains?
For the tool comparison, we select function start detection tools listed in \Cref{tab:approaches} that support PE and have been published within the past decade (2015-2025).
This spans the three learning-based approaches Shin et al.~RNN, XDA, and DeepDi~\cite{DeepDi}, the three non-learning-based tools Nucleus~\cite{nucleus_code}, SMDA~\cite{smda_dis,smda_github}, and rev.ng~\cite{revng}, as well as two popular industry tools IDA Pro 7.7~\cite{idapro} and Ghidra 10.0.4~\cite{ghidra}.
For our one output neuron RNN, we train two models separately for x86 and x64, using the full BAP dataset, described in Section~\ref{sec:bwdataset}, spanning PE samples with ground truth, compiled and linked with MS Visual Studio versions 2010 to 2013 and using the optimization levels Od, O1, Ox, and O2.
Similarly, we use the XDA models described in Section~\ref{xda_reproduction}, trained on 10\% of the BAP-64 dataset. 
For DeepDi, we use the provided model, trained on a mixture of PE files from LLVM, SPEC CPU2006, and SPEC CPU2017.
Note that in contrast to the training datasets used by the learning-based tools, the evaluation dataset spans PE samples built with newer (MSVS 2022) or different (LLVM clang) compilers.
This approach allows us to work towards evaluating the generalizability of the learning-based methods.
\begin{figure*}
\includegraphics[width=\textwidth]{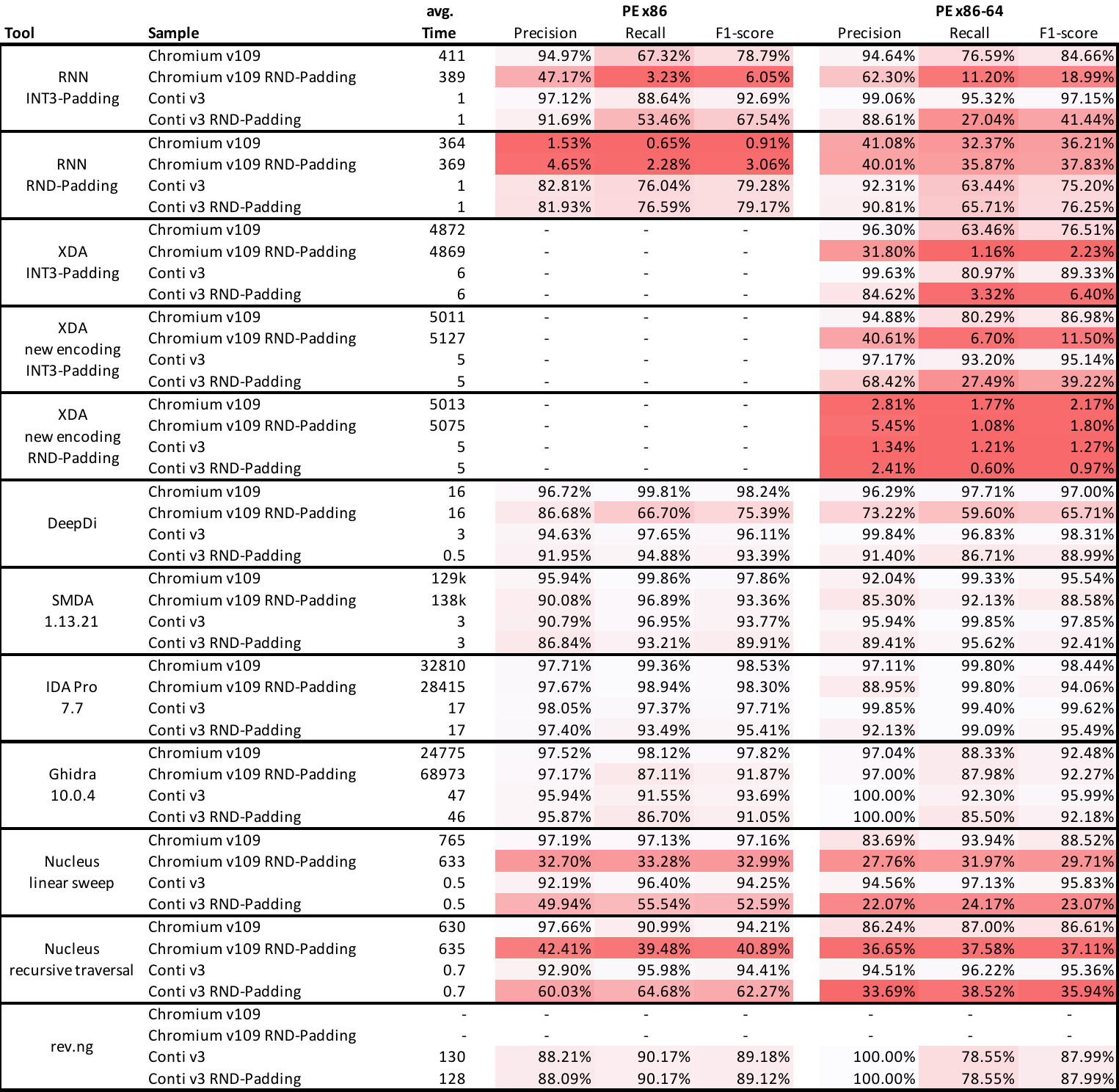}
\caption{Precision, recall, and F\textsubscript{1}-score when predicting function starts in Chromium and Conti samples. Each cell is shaded in red, with the intensity of the red color increasing as the value deviates further from 100\%. \emph{INT3-Padding} in tool names indicates training on samples with \texttt{INT3} instruction as padding between functions, while \emph{RND-Padding} indicates training on samples with random byte values as padding. \emph{RND-padding} in sample names denotes the replacement of compiler-generated padding with random byte values.} \label{tabexp2}
\Description[]{}
\end{figure*}

We evaluated the one output neuron RNN, XDA (in the original and our modified version), DeepDi, SMDA, IDA Pro, Ghidra, Nucleus, and rev.ng on the FuncPEval dataset, specifically Chromium and Conti for both x86 and x64.
The experiments for the RNN, SMDA, IDA, Ghidra, Nucleus, and rev.ng ran on a server with two 2.20 GHz Intel Xeon Silver 4114 CPUs à 20 threads and 128 GB of RAM.
The experiments for XDA and DeepDi ran on a server with two 2.9 GHz AMD EPYC 7542 CPUs à 64 threads, 256 GB of RAM, and one NVIDIA Quadro RTX 8000.

\autoref{tabexp2} shows the processing time for each tool and sample in seconds, averaged over both the x86 and x64 duration.
For Chromium, rev.ng did not finish after 7 days, so we could not obtain results.
For Chromium x64, DeepDi is the fastest, completing in seconds, followed by the RNN and Nucleus, both taking minutes.
XDA, Ghidra, and IDA ran for hours while
SMDA finished after 35 hours.

Concerning F\textsubscript{1}-scores on the x64 Chromium sample, IDA (98.44\%) performs best, followed by DeepDi (97\%), SMDA (95.54\%), and Ghidra (92.48\%).
Nucleus (88.52\%), XDA (86.98\%), and the RNN (84.66\%) score below 90\%.
Even though the F\textsubscript{1}-scores of the RNN and XDA are much lower than in Section~\ref{sec:reproduction}, both still detect a significant number of functions in the evaluation dataset, given the difference between training and evaluation datasets.
Overall, the RNN and XDA exhibit higher precision than recall.
This reflects that most predicted function starts are indeed function starts according to the ground truth, but a significant number of functions is missed (in the worst case, up to 33\% for the RNN and 20\% for XDA).
At first, this might indicate that the tools reliably detect function starts they encountered during training while failing to recognize a significant portion of function starts that were not part of the training data.
However, the results of~\Cref{sec:random_padding} raise doubts whether these tools generalize over code properties observed in function starts.

\begin{lesson}
    IDA, DeepDi, SMDA, and Ghidra achieve the best results in function start detection for PE files, with \fscore{}s exceeding 90\%.
    While Nucleus, XDA, and the RNN produce slightly lower results, with F1-scores above 80\%, they remain valuable tools, particularly due to their relatively fast execution speeds.
    If precise function detection is the primary focus, IDA is the most suitable tool.
    However, if execution speed is a critical factor, DeepDi is the preferred choice, as it delivers the second-best \fscore{}s while being by far the fastest.
\end{lesson}

\subsection{Randomizing the Padding between Functions}
\label{sec:random_padding}

To gain a better understanding of which features are predominantly utilized by the ML-based tools, we analyze the overlap between the detected functions and the training dataset.
The RNN detects a total of 439,363 functions in Chromium x64, with 415,810 of them (95\%) being true positives.
For 353,995 of those functions, we have information about the prologue from the \texttt{.pdata} section.
Only 138,201 out of those 353,955 functions (39\%) have a normalized prologue that also occurs in the training dataset.
Therefore, the RNN detects function starts for which the normalized prologue was not seen during training.
We observe the same for XDA.
Since we do not have training data for DeepDi, we cannot measure the overlap of normalized function prologues.

These values indicate that the RNN and XDA may not primarily rely on function prologues when detecting function starts.
Following Compiler Coding Rule 12 of Intel's Architecture Optimization Reference Manual, compilers align the first instruction of each function at multiples of 16 bytes \cite{guide_intel_opt} and pad the area between the end of the preceding function and the beginning of the next with specific bytes.
Typically, the padding is filled with opcode \texttt{0xcc}, which reflects the mnemonic \texttt{INT 3}, or opcode \texttt{0x90}, which is a single-byte \texttt{NOP} instruction.
Such padding forms a characteristic pattern before a function start.
Therefore, we suspect that the RNN and XDA learn that a series of padding bytes is directly followed by a function start,
a pitfall referred to as a spurious correlation by Arp et al.~\cite{dos_and_donts_machine_learning}, as the padding bytes are not required for the execution of a binary and can be arbitrarily altered in value.
To confirm our hypothesis, we replace the characteristic padding bytes with random byte values in our evaluation dataset and rerun the previous experiment.

\subsubsection{Introducing Random Padding in the FuncPEval Dataset}
\label{sec:exp3dataset}

\begin{figure}
\begin{lstlisting}[language=Python, caption={Pseudocode of algorithm used to replace padding bytes}, label={lst:replace_padding}, basicstyle=\small\ttfamily, breaklines=true]
# iterate over the 20 bytes preceding the function
for i in range(1, 20):
    current_byte =
        sample_bytes[current_function_start - i]
    # make sure the current byte does not belong
    # to another function
    if belongs_to_function(current_byte):
        break
    # make sure the byte is actually a padding
    # byte
    if current_byte.value == 0xcc:
        # replace byte value by random value
        current_byte.value = random_byte()
\end{lstlisting}
\Description[]{}
\end{figure}

We use the FuncPEval dataset described in ~\Cref{sec:evaldataset} and modify the samples to understand the impact of padding bytes on the evaluation.
In our original dataset, we only observe padding with opcode \texttt{0xcc} (mnemonic \texttt{INT 3} representing a software interrupt).
\Cref{lst:replace_padding} shows the algorithm used to replace the padding bytes.
For each function in the samples, we i)~collect up to 20 bytes before the beginning of the function, ii)~consider only those bytes that do not belong to a preceding function, and iii)~replace each padding byte of value \texttt{0xcc} with a random arbitrary byte value.
We select 20 bytes preceding the function to cover any possible padding, which can be up to 15 bytes in length, with an additional margin for tolerance.
This modification does not impact the runtime behavior of the samples because the padding is not part of the control flow and is never executed.
Therefore, malicious actors could arbitrarily alter the values of the padding bytes as an obfuscation to impede automated analyses.
We have observed non-standard padding used by malicious actors in the wild.
The wineloader\footnote{SHA256 hash value of: \ttfamily\footnotesize\seqsplit{72b92683052e0c813890caf7b4f8bfd331a8b2afc324dd545d46138f677178c4}} malware uses opcode \texttt{0xC3} as inter-function padding, which reflects the mnemonic \texttt{RET}.
While possibly an artifact of a rare compiler or compiler configuration, or resulting from deliberate manipulation, the intentions for using non-standard padding are unclear in this specific case.

\subsubsection{Results}
\label{sec:exp3results}
\autoref{tabexp2} shows that the padding significantly impacts the performance of the RNN and XDA.
In the case of Chromium x86, the F\textsubscript{1}-score dropped from 78.79\% to 6.05\% for the RNN. For our modified version of XDA, the F\textsubscript{1}-score dropped from 86.98\% to 11.50\% for Chromium x64.
The modified padding also negatively affects DeepDi, the third machine learning approach, albeit significantly less (drop from 97\% to 65.71\% in F\textsubscript{1}-score for Chromium x64).
We draw two conclusions: First, if a characteristic padding byte pattern is present during training, the RNN and XDA predominantly rely on such a pattern as a delimiter of functions.
Second, unlike RNN and XDA, DeepDi is less affected, most likely as it operates on the granularity of machine code instructions instead of raw bytes.

Given a learning-based approach, it may be considered unfair to train on samples with unmodified padding and evaluate against randomized padding.
To accommodate, we applied the padding randomization to the samples in the BAP training dataset of the RNN and trained a new model to see if the results improve.
The new model (\textit{RND-Padding} in \autoref{tabexp2}) performs overall worse on samples without random padding in comparison to the original model.
In the case of Chromium x86, the new model even produces worse results for the modified sample.
Therefore, we conclude it is not straightforward to train an RNN model that can handle both normal and randomized padding effectively, which raises the question of whether the RNN can identify function start patterns beyond padding.
We also finetuned XDA using a randomized-padding version of the newly encoded dataset.
Overall, the results were significantly worse for both samples with unmodified padding and samples with randomized padding, even though similar results to those in the experiment in Section~\ref{xda_reproduction} were achieved during validation in training.
Consequently, we were unable to train a model for XDA that is robust against randomized padding.
Due to the training code of DeepDi not being publicly available, we could not retrain a version of DeepDi.

The randomized padding also negatively affects IDA, Ghidra, and SMDA; however, the impact on their F\textsubscript{1}-scores never exceeds 10 percent points.
rev.ng is the only tool that is nearly unaffected by the randomized padding; however, we cannot make a statement regarding its results on Chromium.

Nucleus’ detection of function starts is severely negatively impacted by randomized padding.
Both precision and recall are similarly affected.
In the case of Chromium x86, the F1-score drops from 97.16\% to 32.99\%.
Theoretically, modifying padding bytes should not affect Nucleus’ function start detection, unlike the machine learning tools that have learned with padding during training, since Nucleus operates on the interprocedural control flow graph (ICFG), which does not include padding.
Through code review and debugging, we have confirmed that Nucleus is already affected by the randomized padding before creating the ICFG, specifically during disassembling.
Nucleus disassembles using linear sweep.
The randomized padding bytes are interpreted as instructions by the linear sweep disassembler.
This can result in instructions consuming parts of the randomized padding and the beginning of the subsequent function.
In this case, at least the function's first instruction is incorrectly disassembled.
The effect is exacerbated by Nucleus detecting that callers point to the middle of an instruction.
Nucleus attempts to fix this by shifting the start of the basic block to the beginning of the next instruction.
This shift does not resolve the issue, resulting in an incorrect function start being assumed in such cases.
Nucleus also provides an experimental recursive traversal disassembling strategy, which shows no significant improvement.
While the padding bytes are initially ignored by the recursive strategy, a heuristic\footnote{\url{https://bitbucket.org/vusec/nucleus/src/e3ab49db579adbdd8451171e980e9b8f8a546a3c/strategy.cc\#lines-149}} later reconsiders them, leading to the same problem as with the linear sweep.
This heuristic is intended to improve code coverage by assuming another basic block after the end of the current basic block.
Removing this heuristic significantly improves precision but also greatly reduces recall, as the overall code coverage becomes very low.

\begin{lesson}
Function start detection tools are significantly affected by modifications to the compiler-generated padding between functions. When this padding is replaced with random byte values, detection performance deteriorates for the RNN, XDA, DeepDi, and Nucleus. In contrast, IDA, Ghidra, and SMDA are much less impacted. Attempts to retrain the RNN and XDA using samples with randomized padding did not resolve the issue, indicating generalization limitations and leaving it unclear whether these tools can operate effectively on such samples.
\end{lesson}
\section{Limitations \& Conclusion}
\label{sec:conclusion}

We conclude that randomized padding between functions in PE binaries significantly diminishes the effectiveness of the RNN, XDA, and Nucleus.
Even training on samples with randomized padding does not resolve the issue for the learning-based methods RNN and XDA, highlighting their limitations in generalizability.
The remaining tools are also affected by randomized padding, although to a much lesser degree.
Threat actors may evade the affected tools through randomized padding impacting subsequent analysis toolchains, \eg~for malware analysis.
Among the learning-based tools, DeepDi is the least affected and, overall, the fastest.

When considering the unmodified version of Chromium x64, IDA (98.44\%) performs best, closely followed by DeepDi (97\%) and SMDA (95.54\%).
For large-scale applications, DeepDi likely offers the best combination of F1-score and processing speed.

Finally, by modifying the label encoding, we improve XDA’s F\textsubscript{1}-score significantly, resulting in an F\textsubscript{1}-score of 86.98\% for Chromium x64 in comparison to 76.51\% for the unmodified version.

Since no pre-trained models for x86 were provided for XDA, we decided not to include XDA in our x86 evaluation.
Future work could incorporate a newly pre-trained and finetuned version of XDA for x86 in the evaluation.
We assume that the results do not differ significantly between x86 and x64.
Another limitation is that our dataset only contains one malware family (Conti), using a C/C++ code base. 
Ideally, it would be extended to also cover malware using different obfuscation techniques, compilers, and toolchains such as Rust, Nim, Go, and corresponding ground truth.
However, obtaining such ground truth data is challenging due to missing or partial debug symbols.
Finally, while our work focuses on the Windows PE file format, randomized padding likely also impacts the function detection in other executable file formats, such as ELF, given that padding is an architectural recommendation.
Future work could investigate the impact of randomized padding on other file formats.

\section*{Code Artifacts}
To foster future research, we publish our source code, data, and other supplementary information online at: \url{https://github.com/internet-sicherheit/Padding-Matters---Exploring-Function-Detection-in-PE-Files}.

\begin{acks}
The authors gratefully acknowledge funding from the \emph{Federal Ministry of Education and Research} (grants 13FH101KB1 and 16KIS1746), \emph{nicos AG}, and \emph{Cyberus Technology GmbH}.
The authors thank Jan Fedler for his assistance with debugging Nucleus.
\end{acks}

\bibliographystyle{plain}
\bibliography{references}

\end{document}